\documentclass{elsart}

\usepackage{cite}

\begin{document}

\begin{frontmatter}

\title{Iterative solution of differential equations}

\author{Paolo Amore \thanksref{PA}}
\address{Facultad de Ciencias, Universidad de Colima, Bernal D\'iaz del
Castillo 340, Colima, Colima, Mexico}

\author{Hakan Ciftci \thanksref{HC}}

\address{Gazi Universitesi, Fen--Edebiyat Fak\"{u}ltesi,
Fizik B\"{o}l\"{u}m\"{u}, 06500 Teknikokullar,
Ankara, Turkey}

\author{Francisco M. Fern\'{a}ndez \thanksref{FMF}}
\address{INIFTA (Conicet, UNLP), Divisi\'on Qu\'imica Te\'orica,
Diag 113 S/N,  Sucursal 4, Casilla de Correo 16,
1900 La Plata, Argentina}

\thanks[PA]{e--mail: paolo@ucol.mx}
\thanks[HC]{e--mail: hciftci@gazi.edu.tr}
\thanks[FMF]{e--mail: fernande@quimica.unlp.edu.ar}

\begin{abstract}
We discuss alternative iteration methods for differential equations. We
provide a convergence proof for exactly solvable examples and show more
convenient formulas for nontrivial problems.
\end{abstract}

\end{frontmatter}

\section{Introduction \label{sec:Intro}}

The asymptotic iteration method (AIM) is an algorithm for the exact and
approximate solution of second--order ordinary differential equations \cite
{CHS03,F04}. It has been applied to a wide variety of problems that include
exact and approximate calculations of eigenvalues and eigenfunctions,
implementation of perturbation theory, nonrelativistic and relativistic
problems, among others \cite{CHS05,CHS05b,F05,B05,B05b,B06,BKB06,BKYD06,AF06}%
.

Almost all those references are devoted to applications of the approach and
little has been done to provide a sound foundation for the AIM. The proofs
given in the first two papers on the subject \cite{CHS03,F04} do not account
for all the properties of the method. Recently, Matamala et al derived a
most interesting connection between the AIM and continued fractions\cite
{MGD06}.

The purpose of this paper is to provide a deeper insight into the AIM. In
Sec.~\ref{sec:dif_eq} we give well known formal solutions to the
differential equation. In Sec.~\ref{sec:AIM} we discuss the AIM and some
conditions for its successful application to exactly solvable and nontrivial
problems. We discuss a simple problem with constant coefficients in order to
illustrate the kind of solutions expected from the AIM. In Sec.~\ref
{sec:It_Ric_met} we propose alternative iterative approaches that we apply
to the same simple problem just mentioned. In Sec.~\ref{sec:Schrodinger} we
apply the AIM to the harmonic and anharmonic oscillators, compare
alternative recurrence relations and the effect of different initial
conditions. Finally we summarize our results and draw conclusions in Sec.~\ref
{sec:conclusions}.

\section{Differential equation \label{sec:dif_eq}}

The purpose of this paper is the exact or approximate solution of
differential equations of the form
\begin{equation}
y^{\prime \prime }(x)=p(x)y^{\prime }(x)+q(x)y(x)  \label{eq:diff_eq}
\end{equation}
where $p(x)$ and $q(x)$ are integrable functions. We can easily obtain a
formal solution to this equation if we factorize it as
\begin{equation}
\left[ \frac{d}{dx}+a(x)\right] \left[ \frac{d}{dx}+b(x)\right] y(x)=0.
\label{eq:factorization}
\end{equation}
On comparing both equations we realize that $a(x)=-p(x)-b(x)$ where $b(x)$
is a solution of the Riccati equation
\begin{equation}
b^{\prime }(x)-b(x)^{2}-p(x)b(x)+q(x)=0.  \label{eq:Riccati_b}
\end{equation}
Straightforward integration of equation (\ref{eq:factorization}) yields the
general solution
\begin{equation}
y(x)=\exp \left[ -\int^{x}b(u)\,du\right] \left\{ C_{1}+C_{2}\int^{x}\exp
\left[ \int^{t}\left( 2b(u)+p(u)\right) du\right] dt\right\}
\label{eq:gen_sol}
\end{equation}
where $C_{1}$ and $C_{2}$ are integration constants. Notice that it is a
linear combination of two independent solutions of Eq. (\ref{eq:diff_eq}).

The logarithmic derivative
\begin{equation}
f(x)=-\frac{y^{\prime }(x)}{y(x)}  \label{eq:log_der}
\end{equation}
also satisfies the Riccati equation (\ref{eq:Riccati_b}):
\begin{equation}
f^{\prime }(x)-f(x)^{2}-p(x)f(x)+q(x)=0.  \label{eq:Riccati_f}
\end{equation}

All those results are well known, and we summarize them here merely to
facilitate the discussion below.

\section{Asymptotic Iteration Method \label{sec:AIM}}

The AIM is applicable if $p(x)$ and $q(x)$ are $C^{\infty }$ functions. If
we differentiate Eq. (\ref{eq:diff_eq}) $n$ times we obtain
\begin{equation}
y^{(n+2)}(x)=p_{n}(x)y^{\prime }(x)+q_{n}y(x)  \label{eq:y(n+2)}
\end{equation}
where
\begin{eqnarray}
p_{n} &=&p_{n-1}^{\prime }+pp_{n-1}+q_{n-1},  \nonumber \\
q_{n} &=&q_{n-1}^{\prime }+qp_{n-1},\;n=1,2,\ldots  \nonumber \\
p_{0} &=&p,\;q_{0}=q.  \label{eq:pn,qn}
\end{eqnarray}

Ciftci et al \cite{CHS03} proved that if
\begin{equation}
\frac{q_{n}}{p_{n}}=\frac{q_{n-1}}{p_{n-1}}=\alpha  \label{eq:qn/pn}
\end{equation}
then
\begin{equation}
y(x)=\exp \left[ -\int^{x}\alpha (u)\,du\right] \left\{
C_{1}+C_{2}\int^{x}\exp \left[ \int^{t}\left( 2\alpha (u)+p(u)\right)
du\right] dt\right\} .  \label{eq:gen_sol2}
\end{equation}
is a general solution of the differential equation (\ref{eq:diff_eq}). It is
clear that Eq. (\ref{eq:gen_sol2}) agrees with Eq. (\ref{eq:gen_sol}) if $%
\alpha (x)=b(x)$\cite{F04} and its existence is therefore independent of the
condition (\ref{eq:qn/pn}). Moreover, we conclude that $\alpha (x)$ should
satisfy the Riccati equation (\ref{eq:Riccati_b}).

Saad et al\cite{SHC06} proved that the exact solutions just discussed are
polynomial functions. In what follows we provide an alternative proof that
is more convenient for the treatment of nontrivial problems. If we
differentiate the ratio $q_{n-1}/p_{n-1}$ and use the recurrence relations (%
\ref{eq:pn,qn}) we obtain
\begin{equation}
\left( \frac{q_{n-1}}{p_{n-1}}\right) ^{\prime }-\left( \frac{q_{n-1}}{%
p_{n-1}}\right) ^{2}-p\frac{q_{n-1}}{p_{n-1}}+q=\frac{\delta _{n}}{%
p_{n-1}^{2}}.  \label{eq:Ric_qn/pn}
\end{equation}
where
\begin{equation}
\delta _{n}=q_{n}p_{n-1}-q_{n-1}p_{n}.  \label{eq:delta_n(p,q)}
\end{equation}
Therefore, if $\delta _{n}=0$ and $p_{n-1}\neq 0$ then $\alpha
=q_{n-1}/p_{n-1}$ satisfies the Riccati equation (\ref{eq:Riccati_b}) and $%
y=C\exp \left[ -\int^{x}\alpha (u)\,du\right] $ is a solution to the
differential equation (\ref{eq:diff_eq}). Since $y^{(n+1)}=p_{n-1}(y^{^{%
\prime }}+\alpha y)=0$ we conclude that $y(x)$ is polynomial of degree at
most $n$ . Conversely, if $y(x)$ is a polynomial solution of degree $n$,
then $y^{\prime }+q_{n-1}y/p_{n-1}=0$, provided that $p_{n-1}\neq 0$, $%
\alpha =q_{n-1}/p_{n-1}$ satisfies the Riccati equation and $\delta _{n}=0$.
Summarizing, there is a polynomial solution $y(x)$ to Eq. (\ref{eq:diff_eq})
if and only if $\delta _{n}=0$ and $p_{n-1}\neq 0$. This is exactly theorem
2 of reference\cite{SHC06} except that present proof does do not require
that $p_{n}\neq 0$. One can easily verify that if $p_{n}=0$ then $%
q_{k}=p_{k}=0$ for all $k\geq n$ under the conditions above.

The condition (\ref{eq:qn/pn}), although useful for exactly solvable
problems, is not suitable for nontrivial ones where we require that
\begin{equation}
\lim_{n\rightarrow \infty }\frac{q_{n}}{p_{n}}=\alpha .  \label{eq:lim_qn/pn}
\end{equation}
Notice that if
\begin{equation}
\lim_{n\rightarrow \infty }\frac{\delta _{n}}{p_{n-1}^{2}}=0
\label{eq:lim_cond}
\end{equation}
then $\alpha (x)$ given by Eq. (\ref{eq:lim_qn/pn}) is a solution of the
Riccati equation (\ref{eq:Riccati_b}) and Eq. (\ref{eq:gen_sol2}) is the
general solution of the differential equation (\ref{eq:diff_eq}).

In order to understand some of the main features of the AIM it is convenient
to discuss a simple problem already considered earlier. If $p(x)$ and $q(x)$
are constant, then the AIM recurrence relations (\ref{eq:pn,qn}) are exactly
solvable:
\begin{eqnarray}
p_{n} &=&C_{1}\rho _{1}^{n}+C_{2}\rho _{2}^{n},  \nonumber \\
q_{n} &=&qp_{n-1},  \nonumber \\
\rho _{1,2} &=&\frac{p\pm \Delta }{2},\;\Delta =\sqrt{p^{2}+4q},
\label{eq:pn,qn_examp}
\end{eqnarray}
where the constants $C_{1}$ and $C_{2}$ are determined by the conditions $%
p_{0}=p$ and $p_{-1}=1$.

If $|\rho _{1}|<|\rho _{2}|$ equation (\ref{eq:lim_qn/pn}) gives us $\alpha
=-\rho _{1}$ that is a root of the Riccati equation (\ref{eq:Riccati_b}) ($%
b^{\prime }(x)=0$).\ We appreciate that the AIM yields the root with smaller
modulus and we then obtain the general solution by means of Eq. (\ref
{eq:gen_sol2}). This extremely simple exactly solvable example is
interesting because its solutions are not polynomials.

If $|\rho _{1}|=|\rho _{2}|$ and $\rho _{1}\neq \rho _{2}$ the AIM does not
converge but we can overcome this difficulty quite easily. The function $%
v(x)=y(x)\exp (\beta x)$ is a solution of the differential equation $%
v^{\prime \prime }(x)=\tilde{p}v^{\prime }(x)+\tilde{q}v(x)$, where $\tilde{p%
}=2\beta +p$ and $\tilde{q}=q-p\beta -\beta ^{2}$. The new roots are $\tilde{%
\rho}_{1,2}=\beta +\rho _{1,2}$ and the AIM converges for the modified
differential equation because $|\tilde{\rho}_{1}|\neq |\tilde{\rho}_{2}|$.
We will see that transformations of this sort are useful for the treatment
of the Schr\"{o}dinger equation. Clearly, this strategy fails when $\rho
_{1}=\rho _{2}$.

\section{Iterative Riccati method \label{sec:It_Ric_met}}

It is clear from the discussion above that the Riccati equation (\ref
{eq:Riccati_f}) is central to the AIM. We can derive the AIM from the
Riccati equation if we look for a solution of the form
\begin{equation}
f(x)=\frac{A(x)}{B(x)}.  \label{eq:f=A/B}
\end{equation}
On substituting this expression into the Riccati equation and rearranging
conveniently we obtain\cite{F04}
\begin{equation}
\frac{A}{B}=\frac{A^{\prime }+qB}{B^{\prime }+A+pB}.  \label{eq:A/B}
\end{equation}
If we solve the equations $A=A^{\prime }+qB$, and $B=B^{\prime }+A+pB$
iteratively we obtain the AIM recurrence relations
\begin{eqnarray}
A_{n} &=&A_{n-1}^{\prime }+qB_{n-1},  \nonumber \\
B_{n} &=&B_{n-1}^{\prime }+A_{n-1}+pB_{n-1},  \label{eq:An,Bn}
\end{eqnarray}
except that we do not have any prescription for the initial conditions $A_{0}
$ and $B_{0}$. However, according to the example of the preceding section,
the initial conditions do not appear to be that relevant for nonpolynomial
solutions. The most important fact is that
\begin{equation}
\left( \frac{A_{n-1}}{B_{n-1}}\right) ^{\prime }-\left( \frac{A_{n-1}}{%
B_{n-1}}\right) ^{2}-p\frac{A_{n-1}}{B_{n-1}}+q=\frac{\delta _{n}}{%
B_{n-1}^{2}}  \label{eq:Ric_An/Bn}
\end{equation}
where $\delta _{n}=B_{n-1}A_{n}-B_{n}A_{n-1}$, which clearly tell us that
the sequence of ratios $A_{n-1}/B_{n-1}$ may converge to a solution of the
Riccati equation.

We can rearrange the equation for $A$ and $B$ in a different way
\begin{equation}
\frac{A}{B}=\frac{A^{\prime }-pA+qB}{B^{\prime }+A}  \label{eq:A/B_2}
\end{equation}
and derive the alternative recurrence relations

\begin{eqnarray}
A_{n} &=&A_{n-1}^{\prime }-pA_{n-1}+qB_{n-1}  \nonumber \\
B_{n} &=&B_{n-1}^{\prime }+A_{n-1}.  \label{eq:An,Bn_2}
\end{eqnarray}
This approach does not agree with the AIM. We appreciate that the iterative
Riccati method is somewhat more arbitrary than the AIM.

The sequences $\{A_{n}^{(1)},B_{n}^{(1)}\}$ and $\{A_{n}^{(2)},B_{n}^{(2)}\}$
given by the recurrence relations (\ref{eq:An,Bn}) and (\ref{eq:An,Bn_2}),
respectively, are not independent. In fact, it is not difficult to prove
that $A_{n}^{(2)}(x)=A_{n}^{(1)}(x)e^{u(x)}$, and $%
B_{n}^{(2)}(x)=B_{n}^{(1)}(x)e^{u(x)}$, where $u^{\prime }(x)=p(x)$. Both
recurrence relations should give the same result if the initial conditions
are also related by the same transformation.

If we apply equations (\ref{eq:An,Bn}) to the exactly solvable example
discussed above we obtain the AIM result. On the other hand, if we apply
equations (\ref{eq:An,Bn_2}) the result is the root of the Riccati equation
with greater modulus: $\lim_{n\rightarrow \infty }(A_{n}/B_{n})=\rho _{2}$.
In both cases we assume $A_{0}^{\prime }=B_{0}^{\prime }=0$.

\section{The Schr\"{o}dinger equation \label{sec:Schrodinger}}

Direct application of the AIM to the Schr\"{o}dinger equation
\begin{equation}
\psi ^{\prime \prime }(x)=\left[ V(x)-E\right] \psi (x)
\label{eq:Schrodinger}
\end{equation}
may lead to divergent sequences for nonpolynomial solutions. Notice that we
meet the same difficulty when we apply the AIM to the simple example above
with $p=0$. In order to overcome it we make the transformation $\psi
(x)=g(x)y(x)$ and apply the AIM to the resulting differential equation for $%
y(x)$:
\begin{equation}
y^{\prime \prime }=-2\frac{g^{\prime }}{g}y^{\prime }+\left( V-E-\frac{%
g^{\prime \prime }}{g}\right) y.  \label{eq:difeq_y_Schr}
\end{equation}
It has been shown that the rate of convergence of the AIM sequences depends
on the function $g(x)$\cite{F04}. For example, in the case of the harmonic
oscillator $V(x)=x^{2}$ we choose $g(x)=\exp (-x^{2}/2)$ and $y(x)$
satisfies the Hermite differential equation that we discuss briefly below.
Notice that in the case of eigenvalue problems one has to determine the
value of the energy $E$ together with the solution $y(x)$. We obtain the
eigenvalues $E$ from the roots of $\delta _{n}=0$. In the case of polynomial
solutions this equation yields exact eigenvalues for finite $n$, but for
nontrivial problems we obtain increasingly accurate results as $n\rightarrow
\infty $. \cite{CHS03,F04}

In order to have a deeper insight into the approaches derived above it is
convenient to consider a simple differential equation with variable
coefficients $p(x)$ and $q(x)$. One of the simplest examples is the Hermite
differential equation\cite{AS72}
\begin{equation}
y^{\prime \prime }(x)=2xy^{\prime }(x)-2my(x)  \label{eq:Hermite}
\end{equation}
with polynomial solutions for $m=0,1,\ldots $. This example is different
from the preceding one in that the iteration method determines the value of $%
m$ and, consequently, of $q$ together with the solutions of the
corresponding Riccati equation.

The terminating condition yields the values of $m$ corresponding to the
Hermite polynomials as shown by:
\begin{eqnarray}
\delta _{n} &=&-2^{n+1}m(m-1)\ldots (m-n)=0.  \nonumber \\
n &=&1,2,\ldots  \label{eq:delta_Hermite}
\end{eqnarray}
Besides, each of the functions
\begin{eqnarray}
m &=&0\Rightarrow \frac{A_{n}}{B_{n}}=0,\;n\geq 0  \nonumber \\
m &=&1\Rightarrow \frac{A_{n}}{B_{n}}=-\frac{1}{x},\;n\geq 0  \nonumber \\
m &=&2\Rightarrow \frac{A_{n}}{B_{n}}=-\frac{4x}{2x^{2}-1},\;n\geq 1
\nonumber \\
m &=&3\Rightarrow \frac{A_{n}}{B_{n}}=-\frac{3(2x^{2}-1)}{x(2x^{2}-3)}%
,\;n\geq 2  \nonumber \\
&&\cdots  \label{eq:An/Bn_Hermite}
\end{eqnarray}
satisfies the Riccati equation with the corresponding value of $q=-2m$. In
this case we have chosen the AIM initial conditions $A_{0}=q$, and $B_{0}=p$%
. Notice that given the value of $m$ we obtain the exact polynomial solution
from $A_{n}/B_{n}$ for any $n\geq m-1$.

If, on the other hand, we choose, for example, $A_{0}=B_{0}=1$, then, for a
given $m$ we obtain the corresponding Hermite polynomial for all $n\geq m+1$%
. That is to say, we need more iterations for the same result, which
suggests that the AIM prescription is most convenient for this case.

As suggested by the simple example in Sec.~\ref{sec:AIM} the alternative
recurrence relation (\ref{eq:An,Bn_2}) may yield other kind of solutions. We
have confirmed this point in the case of the Hermite equation. The roots of $%
\delta _{n}=0$ are negative integers $m=-1,-2,\ldots $, and $%
A_{n}/B_{n}=-2x,\;-(1+2x^{2})/x,\ldots $ for sufficiently large but finite $%
n $ are solutions to the corresponding Riccati equation.

However, if we choose the initial conditions $A_{0}(x)=q(x)e^{x^{2}}$, and $%
B_{0}=p(x)e^{x^{2}}$ in the recurrence relations (\ref{eq:An,Bn_2}), then we
obtain exactly the AIM results in complete accordance with the discussion in
Sec. \ref{sec:It_Ric_met}.

As a nontrivial model we consider the anharmonic oscillator $V(x)=x^{4}$. We
follow an earlier application of the AIM and choose $g(x)=\exp (\beta
x^{2}/2)$ where $\beta $ is an adjustable parameter\cite{F04}. However, in
this case we select the initial conditions $A_{0}=1$ and $B_{0}=1$
arbitrarily for the recurrence relations (\ref{eq:An,Bn}).

Results are similar to those given by the standard AIM\cite{F04} which shows
that the initial conditions are not so relevant in the case of nonpolynomial
problems. Particularly, if a great number of iterations is required as in
the present application.

\section{Conclusions \label{sec:conclusions}}

In this paper we try to provide an alternative proof for the AIM in the case
of exactly solvable examples and develop equations that appear to be more
convenient for the discussion of nontrivial problems. We also expect to
place the AIM in a more general context of iterative algorithms for
differential equations as Matamala et al have also done regarding the
continued fractions algorithm \cite{MGD06}.

Present results suggest that the AIM gives a convenient prescription for the
starting point of the recurrence relations if one is looking for the square
integrable solutions of the Schr\"{o}dinger equation. However, other initial
conditions may lead to identical results, particularly in the case of
nonpolynomial problems where a great number of iterations is necessary for
accurate results.

One can derive alternative recurrence relations that also give solutions to
the Riccati equation and, consequently, to the linear differential equation.
However, those alternative recurrence relations require that one chooses the
initial conditions carefully; otherwise one may obtain unwanted solutions.

\end{document}